\input{aipcheck}
\documentclass[final]{aipproc}
\layoutstyle{6x9}
\def\be{\begin{equation}}
\def\ee{\end{equation}}
\def\bea{\begin{eqnarray}}
\def\eea{\end{eqnarray}}

\begin{document}
\title[]{Strangeness in the Nucleon}
\author{Harleen Dahiya}{address={Department of Physics (Centre of Advanced Study in
Physics), \\ Panjab University, Chandigarh-160 014, India}}
\author{Manmohan Gupta}{
  address={Department of Physics (Centre of Advanced Study in
Physics), \\ Panjab University, Chandigarh-160 014, India}}

\date{\today}

\begin{abstract}

There are several different experimental indications, such as the
$\sigma_{\pi N}$ term, strange spin polarization, strangeness
contribution to the magnetic moment of the proton, ratio of
strange and non strange quark flavor distributions which suggest
that the nucleon contains a hidden strangeness component which is
contradictory to the naive constituent quark model. Chiral
constituent quark model with configuration mixing
($\chi$CQM$_{{\rm config}}$) is known to provide a satisfactory
explanation of the ``proton spin problem'' and related issues. In
the present work, we have extended the model to carry out the
calculations for the parameters pertaining to the strange quark
content of the nucleon, for example, the strange spin polarization
$\Delta s$, strange components of the weak axial vector form
factors $\Delta \Sigma$ and $\Delta_8$ as well as $F$ and $D$,
strangeness magnetic moment of the proton $\mu_p^s$, the strange
quark content in the nucleon $f_s$ coming from the $\sigma_{\pi
N}$ term, the ratios between strange and non-strange quarks
$\frac{2 s}{u+d}$ and $\frac{2 s}{\bar u+ \bar d}$, contribution
of strangeness to angular momentum sum rule etc.. Our result
demonstrates the broad consistency  with the experimental
observations as well as other theoretical considerations.

\end{abstract}

\maketitle

There is currently enormous interest in the determination of the
strangeness content of the nucleon. It is crucial to our
understanding of Quantum Choromodymnamics (QCD) in the confining
regime and to determine precisely the role played by non-valence
quark flavors in understanding the internal structure of the
nucleon. Theoretically, strange quarks are interesting because
they do not appear explicitly in most quark model descriptions of
the nucleon. The naive constituent quark model \cite{dgg} provides
a useful intuitive picture of the nucleon substructure and has
seen considerable success in accounting for a wide range of
properties of the low-lying hadrons, however one knows that there
is more to the nucleon than the three constituent quarks. A clear
indication in this regard was provided by the EMC measurements of
polarized structure functions of proton in the deep inelastic
scattering (DIS) experiments \cite{emc,adams}, indicating that the
valence quarks of the proton carry only about 30\% of its spin
which includes the contribution even of the strange quark
polarization. Several interesting facts have also been revealed
regarding the quark distribution functions in the DIS experiments
\cite{nmc,na51,e866,hermes} and there are fairly strong signals
indicating that the flavor structure of the nucleon is not limited
to $u$ and $d$ quarks only.

Apart from the indications of DIS data regarding $\Delta s$
\cite{adams} explaining the violation of Ellis Jafe sum rule, this
point is further illustrated by the phenomenological results of
the pion-nucleon sigma term ($\sigma_{\pi N}$) \cite{sigma} which
is extracted from the $\pi N$ scattering data and is a measure of
explicit chiral symmetry breaking in QCD. It gives a strong
indication regarding the strange quark content of proton defined
as$ f_s=\frac{\bar s +s}{\sum (\bar q +q)}$. The OZI rule would
imply $f_s=0$ \cite{cheng}. However, the observed result for
$\sigma_{\pi N}$ indicates that the strange flavor is also present
in the nucleon. Recently, there has been a considerable interest
in calculating the strangeness contribution to the magnetic moment
of the proton $\mu_p^s$  as the same has been measured in the
experiments performed with parity violating elastic
electron-proton scattering at JLab (HAPPEX) \cite{happex} and
MIT-Bates (SAMPLE) \cite{sample}. Similarly, DIS experiments have
given fairly good deal of information regarding the other relavant
observables related to the strange quark content of the nucleon,
for example, the ratios between strange and non-strange quarks
$\frac{2 s}{u+d}$ and $\frac{2 s}{\bar u+ \bar d}$ as measured by
the CCFR Collaboration in their neutrino charm production
experiments \cite{ccfr}.

The chiral constituent quark model ($\chi$CQM), as formulated by
Manohar and Georgi and later developed by Eichten {\it et al.}
\cite{manohar}, can yield an adequate description of the observed
proton flavor and spin structure which is puzzling from the point
of view of naive constituent quark model \cite{cheng}. Further,
chiral constituent quark model with configuration mixing
($\chi$CQM$_{{\rm config}}$) is known to improve the predictions
of $\chi$CQM \cite{hd}. The key to understand the ``proton spin
problem'', in the $\chi$CQM formalism \cite{cheng}, is the
fluctuation process $ q^{\pm} \rightarrow {\rm GB}
  + q^{' \mp} \rightarrow  (q \bar q^{'})
  +q^{'\mp}$, where GB represents the Goldstone boson and $q \bar q^{'}  +q^{'}$
 constitute the ``quark sea'' \cite{cheng,hd,johan}.
The effective Lagrangian describing interaction between quarks and
a nonet of GBs, consisting of octet and a singlet, can be
expressed as $ {\cal L}= g_8 {\bf \bar q}\Phi {\bf q} + g_1{\bf
\bar q}\frac{\eta'}{\sqrt 3}{\bf q}= g_8 {\bf \bar
q}\left(\Phi+\zeta\frac{\eta'}{\sqrt 3}I \right) {\bf q}=g_8 {\bf
\bar q}\left(\Phi'\right) {\bf q}$, where $\zeta=g_1/g_8$, $g_1$
and $g_8$ are the coupling constants for the singlet and octet
GBs, respectively, $I$ is the $3\times 3$ identity matrix. The GB
field $\Phi'$ includes the octet and the singlet GBs. The
parameter $a(=|g_8|^2$) denotes the probability of chiral
fluctuation  $u(d) \rightarrow d(u) + \pi^{+(-)}$, whereas
$\alpha^2 a$, $\beta^2 a$ and $\zeta^2 a$ respectively denote the
probabilities of fluctuations $u(d) \rightarrow s + K^{-(0)}$,
$u(d,s) \rightarrow u(d,s) + \eta$,
 and $u(d,s) \rightarrow u(d,s) + \eta^{'}$.

It would be interesting to mention here that the presence of $s
\bar s$ is not suppressed by the basic mechanism that generates
quark sea. Contribution of the strange quark to the nucleon spin
is one of the major interests in connection with the "Proton Spin
Problem". It is crucial to our understanding of QCD in the
confining regime and gives a direct insight to determine precisely
the role played by heavier, non-valence flavors in understanding
the nucleon internal structure. Almost no information exists,
however, regarding the low-energy manifestations of the sea.
Therefore, it would be interesting to extend the $\chi$CQM$_{{\rm
config}}$ for the calculation of parameters pertaining to the
strangeness content of the nucleon. In particular, we would like
to calculate the strange spin polarization $\Delta s$, strange
components of the weak axial vector form factors $\Delta \Sigma$
and $\Delta_8$ as well as $F$ and $D$, strangeness magnetic moment
of the proton $\mu_p^s$, the strange quark content in the nucleon
$f_s$ coming from the $\sigma_{\pi N}$ term, the ratios between
strange and non-strange quarks $\frac{2 s}{u+d}$ and $\frac{2
s}{\bar u+ \bar d}$, contribution of strangeness to angular
momentum sum rule and the contribution of gluon polarization in
sea. Further, it would also be interesting to carry out a detailed
analysis for the role of SU(3) symmetry breaking and the
strangeness parameters.

To study the role of the strange quarks in the nucleon, one needs
to formulate the experimentally measurable quantities having
implications in this model. The spin structure of a nucleon is
defined as \cite{{cheng},hd,{johan}} $\hat B \equiv \langle
B|N|B\rangle,$ where $|B\rangle$ is the nucleon wavefunction and
$N$ is the number operator giving the number of $q^{\pm}$ quarks.
The contribution to the proton spin in $\chi$CQM$_{{\rm config}}$
is given by the spin polarizations defined as $\Delta q=q^+-q^-$.
After formulating the spin polarizations of various  quarks, we
consider several measured quantities which are expressed in terms
of the above mentioned spin  polarization functions. The
strangeness contribution to the flavor non-singlet components
$\Delta^s_3$ and $\Delta^s_8$, usually calculated in the
$\chi$CQM, are obtained from the neutron $\beta-$decay and the
weak decays of hyperons. The flavor non-singlet component
$\Delta_3$ is related to the well known Bjorken sum rule. Another
quantity which is usually evaluated is the flavor singlet
component $\Delta \Sigma= \frac{1}{2}(\Delta u+\Delta d+\Delta
s)$, in the $\Delta s=0$ limit, this reduces to the Ellis-Jaffe
sum rule.  We have also considered the quark distribution
functions which have implications for the strange quark content.
For example, the antiquark flavor contents of the ``quark sea'',
the strange quark content in the nucleon $f_s$, the ratios between
strange and non-strange quarks $\frac{2 s}{u+d}$ and $\frac{2
s}{\bar u+ \bar d}$. Apart from the above mentioned spin
polarization and quark distribution functions, we have also
calculated the  strangeness magnetic moment of the proton
$\mu_p^s$.

In Table \ref{spin}, we have presented the strangeness parameters
incorporating spin dependent polarization functions along with the
magnetic moments. As is evident from Table \ref{spin}, the
$\chi$CQM$_{{\rm config}}$  is able to give a very good fit for
$\Delta s$ and $\mu_p^s$. It needs to be mentioned that the
strangeness magnetic moment of the proton $\mu_p^s$ is in good
agreement with the HAPPEX data however it is significantly
different when compared with the SAMPLE data, therefore the
quality of numerical agreement can be assessed only after the data
gets refined. It also needs to be mentioned that the strange quark
contribution to the magnetic moment has been subject of intense
experimental and theoretical considerations in the recent times.
The present calculation not only agrees with some theoretical
approaches but is also in agreement with most of the experimental
results. Again, a refinement in the data would tell us about the
extent to which the symmetry breaking values are required.

In Table \ref{quark}, we have presented strange quark flavor
distribution functions. Interestingly, the  $\chi$CQM$_{{\rm
config}}$  is able to give excellent account of the measured
values. The data has been obtained in the case of  $f_s$,
$\frac{2s}{\bar u+\bar d}$, $\frac{2s}{u+d}$,  $\frac{f_3}{f_8}$
wherein we find an almost perfect agreement. Again, refinement of
the data would not only test the $\chi$CQM$_{{\rm config}}$ but
also shed light on the mechanisms of $\chi$CQM$_{{\rm config}}$.
Recently, there has been a lot of interest regarding the parameter
$f_s$, which is related to   $\sigma_{\pi N}$ term obtained from
low energy pion-nucleon scattering. An excellent agreement in the
present case indicates the correct estimation of the role of sea
quarks as has also been advocated by Scadron \cite{scadron}.

In conclusion, it would be interesting to mention that the success
of $\chi$CQM$_{{\rm config}}$ suggests that at leading order, the
model envisages constituent quarks, the octet of Goldstone bosons
($\pi, K, \eta$ mesons) and the {\it weakly} interacting gluons as
appropriate degrees of freedom.

\begin{theacknowledgments}
H.D. would like to thank DST (OYS Scheme), Government of India,
for financial support and the chairman, Department of Physics, for
providing facilities to work in the department.
\end{theacknowledgments}

\begin{table}
\begin{tabular}{|ccccc|} \hline

Parameter & Data  & NRQM & \multicolumn{2}{c|}{$\chi$CQM} \\
\cline{4-5} & & & SU(3) symmetry & SU(3) symmetry breaking \\
\hline

$\Delta s$ & $-0.07 \pm 0.04 $ \cite{adams}   &   0 & $-0.14$&
$-0.03$
\\

$\Delta_8^s$ & $\Delta_8=0.58 \pm 0.025$ \cite{PDG} & 1   & $0.28$
& $0.14$
\\

$\Delta \Sigma^s$ & $\Delta \Sigma=0.31 \pm 0.11$ \cite{PDG}  & 1
& 0.14 & 0.07
\\

$F^s$ & $F=0.462$ \cite{PDG}  & 0.665 &$-0.025$ & $-0.035$
\\

$D^s$ & $D=0.794$\cite{PDG}  & 1 & $0.025$ & 0.035 \\

$\mu_p^s$ & $ -0.038 \pm 0.042$\cite{happex}  & 0 & $-0.06$ &
$-$0.04 \\ & $ -0.36 \pm 0.20$\cite{sample}  &  & & \\ \hline

\end{tabular}
\caption{The calculated values of the strange spin distribution
functions and related parameters.} \label{spin}
\end{table}

\begin{table}
\begin{tabular}{|ccccc|} \hline

Parameter & Data  & NRQM & \multicolumn{2}{c|}{$\chi$CQM} \\
\cline{4-5} & & & SU(3) symmetry & SU(3) symmetry breaking \\
\hline

$\bar s$ & $-$   &   0 & 0.408& $0.11$ \\

$\bar u-\bar d$ & $-0.118 \pm$ 0.015 \cite{e866} & 0   & $-0.118$
& $-0.118$
\\

$\bar u/\bar d$ & 0.67 $\pm$ 0.06 \cite{e866}  & $-$ &0.68 &0.68
\\

$I_G$ & 0.254  $\pm$ 0.005 \cite{e866}  & 0.33 &0.254 & 0.254  \\

$\frac{2 \bar s}{u+d}$ & 0.099$^{+0.009}_{0.006}$ \cite{ccfr} & 0
& 0.236 & 0.09
\\

$\frac{2 \bar s}{\bar u+\bar d}$ & 0.477$^{+0.063}_{0.053}$
\cite{ccfr} & 0 & 1.78 & 0.48
\\

$f_s$ &  0.10 $\pm$ 0.06 \cite{ccfr}  &  0 & 0.18  & 0.09
\\



$f_3/f_8$ & 0.21 $\pm$ 0.05 \cite{ccfr}   &  0.33 &  0.23 & 0.21
\\ \hline

\end{tabular}
\caption{The calculated values of the strange quark flavor
distribution functions and related parameters.} \label{quark}
\end{table}

\end{document}